\documentclass[twocolumn,showpacs,amsmath,amssymb,superscriptaddress]{revtex4}
\usepackage{bm}

\newcommand{\ix}[1]{{\raisebox{0pt}[0pt][0pt]{$\scriptstyle #1$}}}
\newcommand{\eps}{\varepsilon}
\newcommand{\tens}[1]{{\boldsymbol{#1}}}

\newcommand{\xc}{c}
\newcommand{\xco}{C}
\newcommand{\cKY}{h}
\newcommand{\KT}{K}
\newcommand{\grad}{{\tens{d}}}
\newcommand{\eh}{{\tens{e}}}
\newcommand{\eb}{{\tens{e}}}
\newcommand{\noteref}[1]{\cite{#1}}

\DeclareMathOperator{\as}{\mathcal{A}}
\DeclareMathOperator{\tp}{\mathcal{T}}
\DeclareMathOperator{\tr}{\mathrm{tr}}

\begin{document}
\title{Killing--Yano Tensors, Rank-2 Killing Tensors, \\
and Conserved Quantities in Higher Dimensions}

\author{Pavel Krtou\v{s}}

\email{Pavel.Krtous@mff.cuni.cz}

\affiliation{Institute of Theoretical Physics, Charles University, V
Hole\v{s}ovi\v{c}k\'ach 2, Prague, Czech Republic}

\author{David Kubiz\v n\'ak}

\email{kubiznak@phys.ualberta.ca}

\affiliation{Theoretical Physics Institute, University of Alberta, Edmonton,
Alberta, Canada T6G 2G7}

\affiliation{Institute of Theoretical Physics, Charles University, V
Hole\v{s}ovi\v{c}k\'ach 2, Prague, Czech Republic}

\author{Don N. Page}

\email{don@phys.ualberta.ca}

\affiliation{Theoretical Physics Institute, University of Alberta, Edmonton,
Alberta, Canada T6G 2G7}

\author{Valeri P. Frolov}

\email{frolov@phys.ualberta.ca}

\affiliation{Theoretical Physics Institute, University of Alberta, Edmonton,
Alberta, Canada T6G 2G7}

\date{December 4, 2006}

\begin{abstract}

\ From the metric and one Killing--Yano tensor of rank $D-2$ in any
$D$-dimensional spacetime with such a principal Killing--Yano tensor, we show
how to generate $k=[(D+1)/2]$ Killing--Yano tensors, of rank $D-2j$ for all
$0\leq j \leq k-1$, and $k$ rank-2 Killing tensors, giving $k$ constants of
geodesic motion that are in involution.  For the example of the Kerr--NUT--AdS
spacetime (hep-th/0604125) with its principal Killing--Yano tensor
(gr-qc/0610144), these constants and the constants from the $k$ Killing vectors
give $D$ independent constants in involution, making the geodesic motion
completely integrable (hep-th/0611083).  The constants of motion are also
related to the constants recently obtained in the separation of the
Hamilton--Jacobi and Klein--Gordon equations (hep-th/0611245).

\end{abstract}

\pacs{04.70.Bw, 04.50.+h, 04.20.Jb \hfill  Alberta-Thy-16-06}

\maketitle

\section{Introduction}
\label{sc:intro}

In four-dimensional spacetimes like the Kerr metric \cite{Kerr}, the existence
of conserved quantities for geodesics (constants of motion) \cite{Carter} and
the tensorial structures that generate them (Killing vectors, Killing tensors 
\cite{Stac,WP}, and Killing--Yano tensors \cite{Yano,Penrose,Floyd}) have been
very important, not only elucidating particle motion in
these spacetimes, but also leading to the separation of the
Klein--Gordon \cite{Carter}, massless neutrino \cite{Teuk_b, Unruh}, massive
Dirac \cite{Chandrasekhar,Page76}, electromagnetic \cite{Teuk_a}, and
gravitational wave \cite{Teuk_a} equations. 

With the recent interest in higher-dimensional spacetimes, it has become of
interest to extend these old four-dimensional studies to higher dimensions
$D$.  For example, it has been found that the rotating black hole metrics
\cite{MP,HHT,GLPP1,GLPP2,CGLP,CLP} in higher dimensions have a Killing--Yano
tensor of rank $D-2$ \cite{FK,KF} (which we shall call a principal
Killing--Yano tensor) that was used (along with the Killing vectors) to show
\cite{PKVK,KKPV} that geodesic motion in the general $D$-dimensional
Kerr--NUT--AdS rotating black hole spacetime \cite{CLP} is completely
integrable, with $D$ independent constants of motion in involution.

For convenience, we use square brackets to denote the integer part of what is
inside and define $n\equiv [D/2]$, $k\equiv [(D+1)/2]$, and $\varepsilon\equiv
k-n$ (0 for even $D$ and 1 for odd $D$), so
$D=2n+\varepsilon=2k-\varepsilon=k+n$.  Then the Kerr--NUT--AdS spacetimes have
$k$ Killing vectors (giving constants of motion linear in the velocity) and $n$
independent Killing tensors of higher rank (including the metric)
\cite{PKVK,KKPV} (giving other independent constants of motion in involution
that are higher-order polynomials in the velocity).

Here we show how to construct $k$ Killing--Yano tensors, of ranks $D-2j$ for $0
\leq j \leq k-1$, for any spacetime with a principal Killing--Yano tensor. 
Contractions of each of these with itself (leaving two indices free) give $k$
\mbox{rank-2} Killing tensors and hence $k$ independent constants of geodesic
motion in involution for any spacetime with a principal Killing--Yano tensor.

For the case of the Kerr--NUT--AdS spacetimes \cite{CLP}, all of the $k$
Killing vectors can also be constructed from the principal Killing--Yano
tensor  and its covariant derivative, so all $D$ constants of motion arise from
one single Killing--Yano tensor (and the metric, of course, for defining
covariant derivatives and contractions).  For these metrics, we exhibit
explicitly the resulting $k$ Killing--Yano tensors of rank $D-2j$, the $k$
rank-2 Killing tensors, and the $k$ Killing vectors.  (For odd $D$,
$\varepsilon=1$, one of these rank-2 Killing tensors is the tensor product of a
Killing vector with itself and so is not independent or irreducible, leaving
only $D=2k-\varepsilon$ independent rank-2 and rank-1 Killing tensors)  We also
show the relations of the constants of motion arising from all these Killing
tensors with those given in \cite{PKVK,KKPV}, as well as with the constants of
motion arising from the recent separation of the Hamilton--Jacobi and
Klein--Gordon equations \cite{FKK}.

\section{Generating function}
\label{sc:genfc}

The following construction of the Killing tensors, Killing--Yano tensors, and
of the corresponding conserved quantities applies for any metric
\noteref{nt:EuclSign} with a rank-2 closed conformal Killing--Yano tensor
${\tens{\cKY}}$ (or, equivalently, a $(D-2)$-rank Killing--Yano tensor
${\tens{f}=*\,\tens{\cKY}}$; see Section~\ref{sc:KYT}).  Since ${\tens{\cKY}}$
and ${\tens{f}}$ play an important role in our construction, we call both of
them principal tensors.

Let us recall that a rank-2 conformal Killing--Yano tensor ${\tens{\cKY}}$ is
an antisymmetric 2-form which obeys 
\begin{equation}\label{CKYT2}
\nabla_{(a} \cKY_{b)c} = 
    \frac1{D-1}\bigl(g_{ab}\nabla_e\cKY^e{}_c
    -\nabla_e\cKY^e{}_{\!(a}\,g_{b)c}\bigr)\;.
\end{equation}

Assuming the existence of such a tensor,  we define the 2-form
${\tens{F}=\tens{u}\cdot*\,(\tens{u}\cdot*\,\tens{\cKY})}$, which in components
reads
\begin{equation}\label{Fdef}
  F_{ab}=w\, \cKY_{ab}-u_{a}u^{c}\cKY_{cb}-\cKY_{ac}u^{c}u_{b}
  =w\, \cKY_{cd}\, P^c{}_a\, P^d{}_b\;.
\end{equation}
Here ${u^a}$ is the velocity \noteref{nt:velocity},
${w=u_a u^a}$, and ${P^c{}_a=\delta^c{}_a-w^{-1}u^cu_a}$
is the projector to the space orthogonal to ${\tens{u}}$. The 2-form ${\tens{F}}$ is
covariantly conserved in  the direction of ${\tens{u}}$,
\begin{equation}\label{Fcons}
  u^a \nabla_a F_{bc}=0\;.
\end{equation}

Now we introduce the generating function ${W(\beta)}$,
\begin{equation}\label{Wdef}
  W(\beta)=\det\bigl(I+\sqrt{\beta}w^{-1}F\bigr)\;,
\end{equation}
where we take ${F}$ (and similarly ${\cKY}$, ${P}$, ${p}$, and ${Q}$
below) to be the matrix of components ${F^a{}_b}$ of the 2-form ${\tens{F}}$.
Due to the antisymmetry of ${F}$ and properties of the determinant,
${W(\beta)}$ can be rewritten as a function of ${\beta}$ instead of
${\sqrt\beta}$, and in terms of ${\cKY}$ and ${P}$ instead of ${F}$,
\begin{equation}\label{Wdef2}
  W(\beta)
    =\det{}^{\!1/2}\bigl(I-\beta w^{-2}F^2\bigr)
    =\det\bigl(I-\sqrt{\beta}\, \cKY P\bigr)\;.
\end{equation}

Because it is constructed only in terms of 
covariantly conserved quantities ${\tens{F}}$ and ${w}$, 
the generating function is conserved along geodesics,
and the same is true for its derivatives with respect to ${\beta}$.
We can thus define constants of motion ${\xc_j}$ as
the coefficients in the ${\beta}$-expansion of ${W(\beta)}$:
\begin{equation}\label{cdef}
  W(\beta) = \frac1w \,\sum_{j=0}^\infty \xc_j\, \beta^j\;.
\end{equation}
It turns out that all terms with $j>n$ are zero.

To evaluate the observables ${\xc_j}$, we first split ${W(\beta)}$
in the following way:
\begin{equation}\label{Wsep}
  W(\beta) =W_0(\beta)\;\Sigma(\beta)\;,
\end{equation}
with
\begin{equation}\label{WSigmadef}
\begin{aligned}
  &W_0(\beta) = \det\bigl(I-\sqrt{\beta}\cKY\bigr)\;,\\
  &\Sigma(\beta) = \det\Bigl(I\!+\!\frac{\sqrt{\beta}\cKY}{I\!-\!\sqrt{\beta}\cKY}\,p\Bigr) 
  =\tr\bigl(\,(I\!-\!\sqrt{\beta}\cKY)^{\!-\!1}\,p\,\bigr)\;.
\end{aligned}
\end{equation}
Here ${p^a{}_b=w^{-1}u^a u_b}$ is the projector into the ${\tens{u}}$
direction, and we used the fact that the matrix in the determinant  in the
expression for ${\Sigma(\beta)}$ differs from ${I}$ only  in the
one-dimensional subspace given by ${\tens{u}}$.   The generating function thus
splits into a part ${W_0(\beta)}$ independent of ${\tens{u}}$ and a part
${\Sigma(\beta)}$ linear in ${p}$. Using the antisymmetry of ${\cKY}$, we can
rewrite ${W_0(\beta)}$ and ${\Sigma(\beta)}$ in terms of the conformal Killing
tensor with components ${Q^a{}_b=-\cKY^{a}{}_{c}\cKY^{c}{}_{b}}$,
\begin{equation}\label{WSigmainQ}
\begin{aligned}
  W_0(\beta) &= \det{}^{\!1/2}\bigl(I+\beta Q\bigr)\;,\\
  \Sigma(\beta) &= \tr\bigl((I+\beta Q)^{-1}p\bigr) 
           = \sum_{j=0}^{\infty}(-1)^j\tr\bigl(Q^j p)\,\beta^j\;.
\end{aligned}
\end{equation}

We shall assume that ${\cKY}$ is non-degenerate with different 
eigenvalues. This means that there exist ${n}$
uniquely defined 2-dimensional subspaces labeled by the index ${\mu=1,\dots,n}$,
each of which can be spanned by a pair of the orthonormal 
vectors ${\eb_\mu}$ and ${\eh_{\hat\mu}}$ ($\hat \mu\equiv n+\mu$),
in odd number of dimensions complemented with a 
one-dimensional subspace spanned by the vector ${\eh_{\hat0}}$ (${\hat0\equiv2n+1}$),
with non-zero different eigenvalues ${x_\mu}$,
such that ${\tens{\cKY}}$ has the form
\begin{equation}\label{cKYinframe}
  \tens{\cKY}=\sum_{\mu=1}^n x_\mu\,\tens{\omega}^\mu\;.
\end{equation}
Here, ${\tens{\omega}^\mu \equiv \eb^\mu\wedge\eh^{\hat\mu}}$ 
are mutually orthogonal 2-forms
associated with the 2-dimensional planes.
 
Using this frame we can write also the conformal Killing tensor ${\tens{Q}}$ as
\begin{equation}\label{cKTinframe}
  \tens{Q}=\sum_{\mu=1}^n x_\mu^2
    \bigl(\eb^\mu\eb^\mu+\eh^{\hat\mu}\eh^{\hat\mu}\bigr)\;.
\end{equation}

Now we can write down the functions ${W_0}$ and 
${\Sigma}$ in terms of the eigenvalues ${x_\mu}$.
For the part independent of ${\tens{u}}$, we get
\begin{equation}\label{Woinx}
  W_0(\beta)=\prod_{\mu=1}^n(1+\beta x_\mu^2)=\sum_{j=0}^n A^{(j)} \beta^j\;,
\end{equation}
where (cf.\ \cite{CLP}, though here in a more general situation)
\begin{equation}\label{Adef}
  A^{(j)}\equiv\!\!\sum_{\nu_1<\dots<\nu_j}\! x_{\nu_1}^2\dots x_{\nu_j}^2\;.
\end{equation}
Similarly, 
\begin{equation}\label{Sigmainx}
\begin{split}
  \Sigma(\beta)&=\frac1w\Bigl(\eps u_{\hat0}^2+
     \sum_{\mu=1}^n\frac{u_\mu^2+u_{\hat\mu}^2}{1+\beta x_\mu^2}\Bigr)\\
     &=\frac1w\Bigl(\eps u_{\hat0}^2+
     \sum_{j=0}^{\infty}
      (-1)^j\beta^j\sum_{\mu=1}^n(u_\mu^2+u_{\hat\mu}^2)\, x_\mu^{2j}\Bigr)\;,
\end{split}
\end{equation}
with ${u_\mu}$, ${u_{\hat\mu}}$, and ${u_{\hat0}}$ being components of
${\tens{u}}$ with respect to the dual frame
${\eb^\mu,\eh^{\hat\mu},\eh^{\hat0}}$. Recall that ${\eps=0}$ in
even dimensions $D=2n$ and ${\eps=1}$ in odd dimensions $D=2n+1$.

The original generating function reads
\begin{equation}\label{Winw}
  W(\beta) = \frac1w \sum_{j=0}^n 
  \Bigl(\sum_{l=0}^j (-1)^l A^{(j-l)} w_l \Bigr)\beta^j\;,
\end{equation}
where
\begin{equation}\label{wdef}
  w_l = w\tr\bigl(Q^l p\bigr) = \tens{u}\cdot\tens{Q}^{\cdot l}\cdot\tens{u}
   = u_{a_1} Q^{a_1}{}_{\!a_2}\dots Q^{a_l}{}_{\!a}u^a
\end{equation}
are quantities quadratic in the velocity ${\tens{u}}$ given by the $l$-th matrix
power ${\tens{Q}^{\cdot l}}$ of  the conformal Killing tensor ${\tens{Q}}$. 
Clearly, ${w_0=w}$,  and
${w_j=\sum_\mu(u_\mu^2+u_{\hat\mu}^2)x_\mu^{2j}}$ for ${j>0}$. In terms of the
eigenvalues ${x_\mu}$, from the product of Eqs.~\eqref{Woinx} and \eqref{Sigmainx}
we obtain
\begin{equation}\label{Winx}
  W(\beta)=\frac1w\sum_{j=0}^n\Bigl(\eps A^{(j)} u_{\hat0}^2 +
       \sum_{\mu=1}^nA^{(j)}_\mu \bigl(u_\mu^2+u_{\hat\mu}^2\bigr)
        \Bigr)\beta^j\;,
\end{equation}
where we have introduced the quantities (cf.\ \cite{CLP})
\begin{equation}\label{Amudef}
  A^{(j)}_\mu\equiv\!\!
  \sum_{\substack{\nu_1<\dots<\nu_j\\\nu_i\ne\mu}}\!
   x_{\nu_1}^2\dots x_{\nu_j}^2\;.
\end{equation}

\section{Constants of motion and rank-2 Killing tensors}
\label{sc:CKT}

Comparing Eqs.~\eqref{Winw} and \eqref{Winx} with Eq.~\eqref{cdef}, we
can identify the $k = n+\eps =[(D+1)/2]$ conserved quantities ${\xc_j}$
(constants of geodesic motion,  $j=0,\dots,k-1$),
\begin{equation}\label{const}
  \xc_j = \sum_{l=0}^j (-1)^l A^{(j-l)} w_l 
        = \eps A^{(j)} u_{\hat0}^2 + \sum_{\mu=1}^nA^{(j)}_\mu
   \bigl(u_\mu^2+u_{\hat\mu}^2\bigr)\;.
\end{equation}
These constants are quadratic in the velocities. They can be 
generated \cite{WP} by rank-2 Killing tensors  $\tens{\KT}^{(j)}$ as
\begin{equation}\label{constfromKT}
  \xc_j=\KT^{(j)}_{ab} u^a u^b\;,
\end{equation}
where 
\begin{equation}\label{KTdef}
\begin{split}
  \tens{\KT}{}^{(j)} &\equiv \sum_{l=0}^j (-1)^l A^{(j-l)} \tens{Q}^{\cdot l}\\
    &=\eps A^{(j)} \eh^{\hat0}\eh^{\hat0}
      +\sum_{\mu=1}^n A^{(j)}_\mu \bigl(\eb^\mu
       \eb^\mu+\eh^{\hat\mu}\eh^{\hat \mu}\bigr)\;.
\end{split}
\end{equation} 
The matrix power ${\tens{Q}^{\cdot l}}$ of the tensor ${\tens{Q}}$ is
defined in Eq.~\eqref{wdef}.  
The Killing tensors are completely symmetric tensors obeying the equations
\begin{equation}\label{KTe}
\nabla^{}_{\ix{(a}}\KT^{\ix{(j)}}_{\ix{bc)}}=0\;.
\end{equation}

Let us remark that the constant $c_n$ present in an odd number
of spacetime dimensions is the square of the constant corresponding to the
Killing vector $\tens{f}^{(n)}\!$; cf.\ Sections~\ref{sc:KYT} and \ref{sc:KNA}.

The relation \eqref{const} can be inverted using the identities 
\eqref{Ainverse} and \eqref{Aninverse} from the Appendix. We obtain
\begin{equation}\label{vel}
  u_\mu^2+u_{\hat\mu}^2 = 
    U_\mu^{-1}\sum_{j=0}^m (-x_\mu^2)^{n\!-\!1\!-\!j}\; \xc_j\;,
\end{equation}
and, in an odd number of dimensions,
\begin{equation}\label{cn}
  u_{\hat0}^2=\frac{\xc_n}{A^{(n)}}\;.
\end{equation}
Here, the quantity ${U_\mu}$ is defined as (cf.\ \cite{CLP})
\begin{equation}\label{U}
U_{\mu}\equiv\prod_{\nu\ne\mu}(x_{\nu}^2-x_{\mu}^2)\,.
\end{equation}

The coefficients ${A^{(j)}}$ of the ${\beta}$-expansion of ${W_0(\beta)}$ are
the sums of all different products of $j$ different eigenvalues of
${\tens{\cKY}}$, cf.\ Eq.~\eqref{Adef}. Such combinations can be obtained
taking first the $j$-th wedge-power of the 2-form ${\tens{\cKY}}$,
\begin{equation}\label{cKY^k}
\begin{split}
\tens{\cKY}^{\wedge j}\!\!=\tens{\cKY}\wedge\dots\wedge\tens{\cKY}=
  j!\mspace{-15mu}\sum_{\nu_1<\dots<\nu_j}
   \!\!x_{\nu_1}\dots x_{\nu_j}\;\tens{\omega}^{\nu_1}
   \wedge\dots\wedge\tens{\omega}^{\nu_j}\;,
\end{split}\raisetag{2ex}
\end{equation}
and contracting it with itself in all tensor indices. Indeed,
\begin{equation}\label{AisHkHk}
\begin{split}
  A^{(j)}&=\frac{1}{(2j)!(j!)^2}\;
    \tens{\cKY}^{\wedge j}\bullet\tens{\cKY}^{\wedge j}\\
    &=\frac{(2j)!}{(2^{j}j!)^2}\;
    \cKY^{[a_1b_1}\dots \cKY^{a_jb_j]}\cKY_{[a_1b_1}\dots \cKY_{a_jb_j]}\;,
\end{split}
\end{equation}
where ${\bullet}$ denotes the complete contraction, i.e., 
${\tens{B}\bullet\tens{B}}={B_{abc\dots}B^{abc\dots}}$,
and where we used the orthogonality
${\tens{\omega}^\mu\bullet\tens{\omega}^\nu=2\delta^{\mu\nu}}$, along with
the normalization \eqref{wedge}. 

Observing that the relation between the constants ${w^{-1}\xc_j}$ and the
tensor ${w^{-1}\tens{F}}$ is the same as between ${A^{(j)}}$ and
${\tens{\cKY}}$  [cf.\ Eqs.~\eqref{Wdef}, \eqref{cdef}, \eqref{Wsep},
\eqref{WSigmadef}, and \eqref{Woinx}], we obtain a new simple expression for
the constants of motion ${\xc_j}$,
\begin{equation}\label{constinF}
  \xc_j=\frac{1}{(2j)!(j!)^2}\,w^{1-2j}\;\tens{F}^{\wedge j}\bullet
  \tens{F}^{\wedge j}\;.
\end{equation}
If one defines a $1$-form $\tens{v}$ with components 
\begin{equation}\label{vdef}
  v_a = \cKY_{ab}u^b\;,
\end{equation}
orthogonal to the velocity 1-form ${\tens{u}}$, then Eq.~\eqref{Fdef} implies
that
\begin{equation}\label{Frel}
\tens{F} = w \tens{\cKY} + \tens{u} \wedge \tens{v}\;.
\end{equation}
Since $\tens{u} \wedge \tens{v} \wedge \tens{u} \wedge \tens{v} = 0$,
we have
\begin{equation}\label{Fsum}
\tens{F}^{\wedge j} = 
w^j \tens{\cKY}^{\wedge j} 
+ j w^{j-1} \tens{u} \wedge \tens{v}\wedge\tens{\cKY}^{\wedge(j-1)}\;.
\end{equation}
The total contraction of this with itself is
\begin{equation}\label{FkFk}
\begin{split}
&\tens{F}^{\wedge j}\bullet\tens{F}^{\wedge j} = 
 w^{2j} \tens{\cKY}^{\wedge j}\bullet\tens{\cKY}^{\wedge j}\\
 &\quad + 2j w^{2j-1} \tens{\cKY}^{\wedge j}\bullet
 \bigl(\tens{u} \wedge \tens{v}\wedge\tens{\cKY}^{\wedge(j-1)}\bigr)\\
 &\quad + j^2 w^{2j-2} 
 \bigl(\tens{u} \wedge \tens{v}\wedge\tens{\cKY}^{\wedge(j-1)}\bigr)
 \bullet\bigl(\tens{u} \wedge \tens{v}\wedge\tens{\cKY}^{\wedge(j-1)}\bigr)\;.
\end{split}\raisetag{7ex}
\end{equation}

In the total contraction of the last term, $\tens{u}\cdot\tens{v} = 0$, and any
term with a contraction of $\tens{u}$ with any of the $\tens{\cKY}$'s gives
another $\tens{v}$ which combines with the original $\tens{v}$ to give zero by
the antisymmetry of the wedge product.  Therefore, the only nonzero parts of
the last term have another $w = \tens{u}\cdot\tens{u}$ factor, giving a total
factor of $w^{2j-1}$ for the total contraction of that term.  Upon the
substitution of Eq.~\eqref{FkFk} into \eqref{constinF}, the dependence on ${w}$
cancels out, and we recover the quadratic dependence on ${\tens{u}}$ which
enters through ${\tens{u}}$ and ${\tens{v}}$. Comparing with
\eqref{constfromKT}, we can write the tensorial relation between
${\tens{\KT}{}^{(j)}}$ and ${\tens{\cKY}}$, which in components reads
\begin{equation}\label{KTcomp}
\begin{split}
  \KT{}^{(j)}{}^a{}_b 
  &= \frac{(2j)!}{(2^{j}j!)^2}\Bigl(
    \delta^a_b\;\cKY^{[a_1b_1}\dots \cKY^{a_jb_j]}\cKY_{[a_1b_1}\dots \cKY_{a_jb_j]}\\
    &\qquad\quad-4j\,\cKY^{[ab_1}\dots \cKY^{a_jb_j]}\cKY_{b[b_1}\dots \cKY_{a_jb_j]}\\
    &\qquad\quad+2j\,\cKY^{a[b_1}\dots \cKY^{a_jb_j]}\cKY_{b[b_1}\dots \cKY_{a_jb_j]}
    \Bigr)\\
  &=\frac{(2j)!}{(2^{j}j!)^2}\Bigl(
    \delta^a_b\;\cKY^{[a_1b_1}\dots \cKY^{a_jb_j]}\cKY_{[a_1b_1}\dots \cKY_{a_jb_j]}\\
    &\qquad\quad-2j\,\cKY^{[ab_1}\dots \cKY^{a_jb_j]}\cKY_{[bb_1}\dots \cKY_{a_jb_j]}\Bigr)\;,
\end{split}
\end{equation}
where we have employed the definition \eqref{vdef}, the identities \eqref{asid2} and
\eqref{asid3}, and the normalization \eqref{wedge}.

Recently \cite{PKVK} there have been found different conserved quantities,
\begin{equation}\label{oldc}
  \xco_j=w^{-j}\tr\!\left[(-F^2)^j\right]\;,
\end{equation}
which are, however, not quadratic in velocities. Now we show that also these
observables can be generated from the generating function ${W(\beta)}$. 
Taking the logarithm of Eq.~\eqref{Wdef2} and expanding it into a power series,
we obtain
\begin{equation}\label{logW}
  \log W(\beta) = \frac12\tr\log\bigl(I-\beta w^{-2}F^2\bigr)
    = \sum_{j=0}^\infty\frac{(-1)^{j+1}}{2j}\frac{\beta^j}{w^j}\,\xco_j\;.
\end{equation} 
The constants ${\xco_j}$ are thus (up to constant factors and powers of ${w}$)
given by derivatives of ${\log W(\beta)}$. We also obtained 
the relation between both sets of constants
which can be formulated as
\begin{equation}\label{ccrel}
  \sum_{j=0}^{\infty}w^{-1}\xc_j\beta^j = 
  \exp\Bigl(\sum_{j=0}^\infty\frac{(-1)^{j+1}}{2j}
  \frac{\beta^j}{w^j}\xco_j\Bigr)\;.
\end{equation}
Comparing different orders of ${\beta}$ we get for the first four constants
\begin{equation}\label{ccrelex}
\begin{split}
  \xc_1&=-\frac12\xco_1\;,\\
  w\,\xc_2&=-\frac14\xco_2+\frac18\xco_1^2\;,\\
  w^2\xc_3&=-\frac16\xco_3+\frac18\xco_1\xco_2-\frac1{48}\xco_1^3\;,\\
  w^3\xc_4&=-\frac18\xco_3\!+\!\frac1{12}\xco_1\xco_3\!+\!
  \frac1{32}\xco_2^2\!-\!\frac1{32}\xco_1^2\xco_2\!+\!\frac1{384}\xco_1^3\;.
\end{split}\raisetag{14ex}
\end{equation}
It is shown in \cite{KKPV} that the observables ${\xco_j}$ Poisson commute
between each other. The relation \eqref{ccrel}, which shows that the
${\xc_j}$'s are polynomial combinations of the ${\xco_j}$'s and ${w}$ with
constant coefficients, thus proves that also the observables ${\xc_j}$ are in
involution,
\begin{equation}\label{Poisson}
  \{\xc_i,\xc_j\}=0\;.
\end{equation} 
This gives non-trivial relations for the corresponding Killing tensors, namely
\begin{equation}\label{PoissonK}
  \KT{}^{(j)}_{e(a}\, \nabla^{e} \KT{}^{(l)}_{bc)}- 
  \KT{}^{(l)}_{e(a}\, \nabla^{e} \KT{}^{(j)}_{bc)}=0\;.
\end{equation}

\section{Killing--Yano tensors}
\label{sc:KYT}

The existence of the closed conformal Killing--Yano tensor ${\tens{\cKY}}$
guarantees the existence of the Killing--Yano tensor ${\tens{f}}$ which is
obtained by the Hodge dual ${\tens{f}=*\,\tens{\cKY}}$ \cite{Cari}.  This principal 
Killing--Yano tensor enables one to construct  a rank-2 Killing
tensor ${K^a{}_b=f^{ae_1\dots e_{D-3}}f_{be_1\dots e_{D-3}}}$. Here we
demonstrate that all the Killing tensors found in the previous section can be
constructed in a similar way.

First, let us recall that a conformal Killing--Yano tensor (CKYT) 
\cite{Tachi, Kashi, TKa}
of a general rank ${r}$ is an antisymmetric 
${r}$-form ${\tens{f}}$ the covariant derivative
of which can be split into the antisymmetric and divergence parts
\begin{equation}\label{CKYTdef}
  \nabla\tens{f}=\as\nabla\tens{f}+\tp\nabla\tens{f}\;.
\end{equation}
Here ${\as}$ is the standard anti-symmetrization and ${\tp}$ 
is the projection onto the `trace' part of the tensor of rank ${r+1}$ which 
is antisymmetric in the last ${r}$ indices,
\begin{equation}\label{dpdef}
  \tp A_{aa_1\dots a_r} = \frac{r}{D-r+1}\,g_{a[a_1}A^e{}_{|e|a_2\dots a_r]}\;.
\end{equation}

The operation ${\tp}$ satisfies ${\tp^2=\tp}$ and ${\tp\!\as=\as\tp=0}$. This
means that a tensor ${\tens{A}}$ satisfies ${\tp\! \tens{A}=\tens{A}}$  if and
only if it has the form ${A_{aa_1\dots a_r}}={g_{a[a_1}\alpha_{a_2\dots
a_r]}}$. The divergence part ${\tp\nabla\tens{f}}$ thus depends only  on the
divergence ${\nabla_e f^e{}_{ab\dots}}$. The condition \eqref{CKYTdef} 
implies that ${\nabla\tens{f}}$ does not have a harmonic part \cite{Yanobook}
(given by the complement of the ${\as}$ and ${\tp}$ projectors), i.e.,
${\tens{f}}$ does not have a part for which both  ${\grad\,\tens{f}}$ and
${\nabla{\cdot}\tens{f}}$ vanishes. 

A CKYT transforms into a CKYT under Hodge duality. 
The antisymmetric part ${\as\nabla\tens{f}}$ transforms 
into the divergence part ${\tp\nabla{*}\!\tens{f}}$ and vice versa. 

A Killing--Yano tensor ${\tens{f}}$ is such a CKYT for which
the divergence part is missing, i.e.,
\begin{equation}\label{KYTdef}
  \nabla\tens{f}=\as\nabla\tens{f}\;.
\end{equation}
The dual of a Killing--Yano tensor is a closed CKYT (see also \cite{Cari}),
i.e., an ${r}$-form obeying
\begin{equation}\label{CCKYTdef}
  \nabla\tens{f}=\tp\nabla\tens{f}\;.
\end{equation}

The wedge product of two closed CKYTs is again a closed CKYT [see
Eq.~\eqref{wedgeCCKYT} in the Appendix].   We can thus start with the principal
closed CKYT ${\tens{\cKY}}$ and construct its wedge powers
${\tens{\cKY}^{\wedge j}}$ ($j=0,\dots,k-1$), which are again closed
CKYTs. Their duals, 
\begin{equation}\label{KYTfromCKY}
  \tens{f}^{(j)} = *\, \tens{\cKY}^{\wedge j}\;,
\end{equation}
are then Killing--Yano tensors of rank ${D-2j}$. Their components are
\begin{equation}\label{KYTfromCKYcomp}
  f^{(j)}_{a_1\dots a_{D-2j}}
  =2^{-j}\eps_{a_1\dots a_{D\!-2j}}{}^{e_1\dots e_{2j}}
  \cKY_{e_1e_2}\dots \cKY_{e_{2j\!-\!1}e_{2j}}\;,
\end{equation}
where ${\eps_{a_1\dots a_D}}$ are components of the Levi-Civita tensor ${\tens{\eps}}$
and the normalization \eqref{wedge} has been employed.

Now we show that these Killing--Yano tensors generate the rank-2 Killing
tensors ${\tens{\KT}{}^{(j)}}$ constructed above. Namely, using the identity
\eqref{asid1}, we write
\begin{equation}\label{KYsquare}
\begin{split}
  &{\frac{1}{(D\!-\!2j\!-\!1)!(j!)^2}}\;
    f^{\ix{(j)\,ae_1\dots e_{D\!-\!2j}}}_{} f^{\ix{(j)}}_{\ix{be_1\dots e_{D\!-\!2j}}} =\\ 
    &\quad=\frac{(2j\!+\!1)!}{(2^j j!)^2}\; \delta{}^{\ix{[a}}_{\ix{[b}} 
    \cKY^{\ix{a_1b_1}}_{}\dots\cKY^{\ix{a_{j}b_{j}]}_{}}
    \cKY^{}_{\ix{a_1b_2}}\dots\cKY^{}_{\ix{a_{j}b_{j}]}}\;.
\end{split}
\end{equation}
With the help of Eq.~\eqref{asid2}, we see that the last expression coincides
with  the formula \eqref{KTcomp} for the Killing tensors. So we have
\begin{equation}\label{KTKYT}
  \KT{}^{(j)}{}^a{}_b = {\frac{1}{(D\!-\!2j\!-\!1)!(j!)^2}}\;
    f^{(j)\,ae_1\dots e_{D\!-\!2j}} f^{(j)}_{be_1\dots e_{D\!-\!2j}}\;.
\end{equation}

Starting from the principal closed CKYT ${\tens{\cKY}}$, we build  the sequence
of closed CKYTs ${\tens{\cKY}^{\wedge j}}$, given explicitly in
Eq.~\eqref{cKY^k}, which generates Killing--Yano tensors
${\tens{f}^{(j)}=*\,\tens{\cKY}^{\wedge j}}$. These Killing--Yano tensors can
be used to construct the rank-2 Killing tensors ${\tens{\KT}{}^{(j)}}$ given by
the formula \eqref{KTKYT}, or explicitly by Eq.~\eqref{KTdef} or
\eqref{KTcomp}. 

In particular, in an odd number of spacetime dimensions, the last Killing--Yano
tensor  ${\tens{f}^{(n)}}\propto {\sqrt{A^{(n)}}\,\eh_{\hat0}}\,$ is a Killing
vector.  Obviously, the corresponding Killing tensor $\tens{\KT}^{(n)}\propto
{\tens{f}^{(n)}}{\tens{f}^{(n)}}$ is reducible.

\section{Kerr--NUT--AdS spacetimes}
\label{sc:KNA}

We shall now demonstrate that the structure explored above is
fully realized in the Kerr--NUT--AdS spacetimes \cite{CLP}.

In the notation of previous sections
(i.e., using the base of 1-forms 
${\eb^a=\{\eb^\mu,\eh^{\hat\mu},\eh^{\hat0}\}}$, 
${\mu=1,\dots,n}$, ${\hat\mu=\mu+n}$, ${\hat0=2n+1}$,
with the 1-form ${\eh^{\hat0}}$ present only for odd ${D}$),
the Kerr--NUT--AdS metric may be written in the orthonormal form
\begin{equation}\label{metrics}
\tens{g}
     = \sum_{a=1}^D \eb^a \eb^a
     = \sum_{\mu=1}^n 
          (\eb^\mu \eb^\mu + \eh^{\hat \mu} \eh^{\hat \mu})
       + \eps\, \eh^{\hat 0}\eh^{\hat 0} \, ,
\end{equation}
where the orthonormal basis one-forms are
\begin{equation}\label{one-forms}
\begin{aligned}
\eb^\mu &= Q_{\mu}^{-1/2} \grad x_{\mu}\;,\\\
\eh^{\hat \mu} &= Q_{\mu}^{1/2}
 \sum_{j=0}^{n-1}A_{\mu}^{(j)}\grad\psi_j\;,\\
\eh^{\hat 0} &= (-c/A^{(n)})^{1/2}
\sum_{j=0}^nA^{(j)}\grad\psi_j\;.
\end{aligned}
\end{equation}

The quantities $A^{(j)}$,  $A_{\mu}^{(j)}$ in terms of coordinates ${x_\mu}$
are of the form of Eqs.~\eqref{Adef} and \eqref{Amudef}, 
$Q_{\mu}={X_{\mu}}/{U_{\mu}}\,$ with $U_{\mu}$ given by Eq.~\eqref{U},
$c=\prod_{j=1}^{k-1} a_j^2\,,$ and 
\begin{gather}
X_{\mu}=(-1)^{1-\eps}\frac{1+\lambda x_{\mu}^2}{x_{\mu}^{2\eps}}
\prod_{j=1}^{k-1}(a_j^2-x_{\mu}^2)
+2M_{\mu}(-x_{\mu})^{1-\eps}.
\label{QXUcdef}
\end{gather}
The constants $({M_\mu}, a_j)$ are related to the mass, NUT parameters, and
angular momenta, and ${\lambda}$ is  proportional to the cosmological constant
\cite{CLP}.  The metric represents an Einstein space obeying the Einstein
equation 
\begin{equation}\label{Einstein}
{R_{ab}=(D-1)\lambda g_{ab}}\;.
\end{equation}

Using the identities \eqref{Ainverse} and \eqref{Aninverse}, we find the dual
vectors
\begin{equation}\label{basisvectors}
\begin{aligned}
\eb_\mu &= Q_{\mu}^{1/2} \tens{\partial}_{x_\mu}\,,\\
\eh_{\hat\mu}&= \frac{1}{Q_{\mu}^{1/2}U_{\mu}}\,
\sum_{j=0}^{k-1}(-x_{\mu}^2)^{n-1-j}\tens{\partial}_{\psi_{j}}\,,\\
\eh_{\hat 0} &= \bigl(-cA^{(n)}\bigr)^{-1/2}\tens{\partial}_{\psi_n}\,,
\end{aligned}
\end{equation}
and the corresponding inverse relations 
\begin{equation}\label{coorvectors}
\begin{aligned}
\tens{\partial}_{x_\mu} &= Q_{\mu}^{-1/2} \eb_\mu \,,\\
\tens{\partial}_{\psi_j}&=\eps A^{(j)}
  \Bigl(-\frac{c}{A^{(n)}}\Bigr)^{1/2}\eh_{\hat 0}
  +\sum_{\mu=1}^nQ_{\mu}^{1/2}A_{\mu}^{(j)}\eh_{\hat \mu}\,,\\
\tens{\partial}_{\psi_n}&=\bigl(-{c}{A^{(n)}}\bigr)^{1/2}\eh_{\hat 0}\;.
\end{aligned}
\end{equation}

It is possible to prove \noteref{nt:KYproof} that that the $\tens{h}$ found in
\cite{KF} (there called $\tens{k}$) in this metric in all dimensions $D$ is a
principal closed CKYT, which in the frame \eqref{one-forms} takes exactly the
form \eqref{cKYinframe}. This means that the generally defined eigenvalues
$x_{\mu}$  of the principal CKYT  ${\tens{\cKY}}$ \eqref{cKYinframe} coincide
with the chosen (`natural') coordinates ${x_\mu}$ of the Kerr--NUT--AdS metric.

We now demonstrate that from the very existence of this tensor one can extract
all the  constants of geodesic motion for the Kerr--NUT--AdS metrics. Namely,
besides the $k$ constants of motion \eqref{const} connected with the rank-2
Killing tensors \eqref{KTdef}, also all the $k$ isometries follow from the
existence of the principal CKYT ${\tens{\cKY}}$. 

First of all, it was proved in \cite{Jez} that in an Einstein space, obeying
Eq.~\eqref{Einstein}, the divergence $\tens{\xi}$ of a CKYT ${\tens{\cKY}}$,
\begin{equation}
\xi^a=\frac{1}{D-1}\nabla_{\!b}\,\cKY^{ba}\,,
\end{equation}
is a Killing vector. In particular we find ${\tens{\xi}=\tens{\partial}_{\psi_0}}$ \cite{KF}.

Next, using Eqs.~\eqref{KTdef} and \eqref{coorvectors}, 
we can recover $n-1$ other Killing vectors ${\tens{\partial}_{\psi_j}}$,
\begin{equation}\label{etaj}
(\partial_{\psi_j})^a=\KT^{(j)}\,\!^a_{\ b}\xi^b,\quad j=1,\dots,n-1\;. 
\end{equation}
For a similar construction in 4D see \cite{Somm, Coll, Rud}.  Finally, as
mentioned in Section~\ref{sc:KYT},  in odd dimensions the last Killing vector is given by
the ${n}$-th Killing--Yano tensor  $\tens{f}^{(n)}$, which in the present
example turns out to be $\tens{\partial}_{\psi_n}$.

It would be very interesting to find under what general conditions on the CKYT
$\tens{\cKY}$, and possibly on the curvature, this construction gives all the
isometries present in the spacetime.

The constants of geodesic motion in the higher-dimensional Kerr--NUT--AdS
spacetime are now completely determined.  Denoting the constants from the
Killing vectors as
\begin{equation}
b_j=(\partial_{\psi_j})^au_a\,,\qquad j=0,\dots, k-1,
\end{equation}
we first find the frame components of velocity $u_{\hat \mu}$, and
possibly $u_{\hat 0}$.  It follows from Eq.~\eqref{basisvectors} that
\begin{equation}
u_{\hat\mu}=\frac{1}{Q_{\mu}^{1/2}U_{\mu}}\,
\sum_{j=0}^{k-1}(-x_{\mu}^2)^{n-1-j}b_{j}\,,\ 
u_{\hat 0} = \bigl(-cA^{(n)}\bigr)^{-1/2}b_n\,.
\end{equation}
Comparing with (\ref{cn}) we find 
\begin{equation}\label{cnbn}
c_n=-\frac{b_n^2}{c}\;,
\end{equation}
which clearly illustrates the fact that the constant $c_n$ corresponds to the
reducible Killing tensor $\tens{K}^{(n)}\propto\tens{f}^{(n)}\tens{f}^{(n)}$.
The remaining components of velocity, $u_{\mu}$, are given (up to signs) in
terms of the constants $b_j$, $j=0,\dots, k-1$, and $c_{j}$, $j=0,\dots, n-1$
[which correspond to the irreducible Killing tensors (\ref{KTdef})] by
Eq.~(\ref{vel}).

The existence of ${n}$ rank-2 irreducible Killing tensors
${\tens{\KT}{}^{(j)}}$ and $k={D-n}$ Killing vectors
${\tens{\partial}_{\psi_j}}$ is closely related to the question of separability
of the Hamilton--Jacobi  and Klein--Gordon equations. It is shown in
\cite{Benenti, Francaviglia} that if the spacetime possesses such tensors that
satisfy the condition \eqref{PoissonK}, 
${\mathcal{L}_{\tens{\partial}_{\psi_j}}\tens{\KT}{}^{(j)}=0}$ with
${\{\tens{\partial}_{\psi_j},\tens{\partial}_{\psi_l}\}=0}$---which holds in our
case---then there exists a so-called {\em separability structure}. This
structure guarantees the separability of the Hamilton--Jacobi equation and for
Einstein spaces also the separability of the Klein--Gordon equation.  

The separability of these equations was explicitly demonstrated recently
\cite{FKK}. It turns out that the integration constants obtained by the
separation  of the Hamilton--Jacobi equation are the quantities ${\xc_j}$ given
by Eq.~\eqref{const}. Indeed, if we transform the tetrad components 
${u_\mu,u_{\hat\mu}}$ and ${u_{\hat0}}$ into the coordinate frame we find that
the expression ${U_\mu(u_\mu^2+u_{\hat\mu}^2)}$ corresponds to the quantity
${F_\mu}$ of \cite{FKK}. Comparing Eqs.~\eqref{vel} and (14) of
\cite{FKK}, we can identify the constants defined above with those from
\cite{FKK} (cf. also  Eqs.~\eqref{cn} and \eqref{cnbn} with Eq.~(15) of
\cite{FKK}).

\section{Discussion}
\label{sc:disc}

We have seen that the existence of a principal Killing--Yano tensor, one of
rank $D-2$, guarantees the existence of $k=D-n$ Killing--Yano tensors of rank
$(D-2j)$,  $j=0,\dots,k-1$, and that each of these Killing--Yano tensors
generates a Killing tensor of rank $2$, $n$ of which are irreducible. In the
case of the Kerr--NUT--AdS spacetimes, the principal Killing--Yano tensor also
generates all the $k$ Killing vectors, and hence all $D$ of the independent
constants of geodesic motion.

Our results raise various questions.  For example, is the construction of
Killing vectors by a rank-$(D-2)$ Killing--Yano tensor general, or specific to
only certain metrics?  For what classes of spacetimes are there rank-$(D-2)$
Killing--Yano tensors?  For what classes of such spacetimes does the
rank-$(D-2)$ Killing--Yano tensor generate enough Killing tensors to give $D$
independent constants of motion?  Are there any new Einstein metrics within
these classes?  Do these structures enable one to separate the Dirac,
electromagnetic, and gravitational wave equations in the Kerr--NUT--AdS
spacetimes and/or in any other possible members of these classes?
What is the relation of the existence of principal Killing--Yano tensors 
to the algebraic type of the metric?
We suspect that our observations may be the tip of an iceberg
of important new relations for higher dimensional spacetime metrics.

\section*{Acknowledgments}

P.K.\ is supported by  the grant GA\v{C}R 202/06/0041 and 
appreciates the hospitality of the University of Alberta. 
D.K.\ is grateful to the Golden Bell Jar Graduate
Scholarship in Physics at the University of Alberta. 
D.P.\ thanks the Natural Sciences and Engineering 
Research Council of Canada for financial support.
V.F.\ thanks the Natural Sciences and Engineering Research 
Council of Canada and the Killam Trust for financial support. 
We have benefited from discussions with Paul Davis and Gary Gibbons.

\appendix

\section{Useful identities}
\label{sc:apx}

In this Appendix we list some identities used in the main text.
Their proofs are mostly straightforward but also lengthy and cumbersome. 

First we list three identities for antisymmetric tensors. 
The Levi-Civita tensor satisfies  
\begin{equation}
  \label{asid1}
  \eps^{a_1\dots a_rc_{r+1}\dots c_D} \eps_{b_1\dots b_rc_{r+1}\dots c_D}
  = r! (D-r)! \delta^{[a_1}_{b_1}\dots\delta^{a_r]}_{b_r}\;,
\end{equation}
the projector on the antisymmetric tensors can be split as
\begin{equation}
  \label{asid2}
  (r+1)\,\delta^{[a}_{[b}\delta^{a_1}_{b_1}\dots\delta^{a_r]}_{b_r]}=
  \delta^{a}_{b}\delta^{[a_1}_{[b_1}\dots\delta^{a_r]}_{b_r]}
  -r\,\delta^{a}_{[b_1}\delta^{[a_1}_{|b|}\dots\delta^{a_r]}_{b_r]}
\end{equation}
and finally
\begin{equation}\label{asid3}
  \cKY_{[a_1b_1}\dots \cKY_{a_jb_j]}=\cKY_{a_1[b_1}\dots \cKY_{a_jb_j]}\;,
\end{equation}
which holds for any antisymmetric tensor ${\cKY_{ab}}$.

In our convention the wedge product is proportional to 
the anti-symmetrization of the tensor product which for the
${j}$-th wedge-power of a 2-form ${\tens{\cKY}}$ gives
\begin{equation}\label{wedge}
\tens{\cKY}^{\wedge j}=\frac{(2j)!}{(2!)^j}\as\tens{\cKY}^j\;.
\end{equation}

Next we want to show that the wedge product of two 
closed CKYTs ${\tens{p}}$ and ${\tens{q}}$
of rank ${r}$ and ${s}$ respectively is again a closed CKYT.
It is clear that ${\tens{p}\wedge\tens{q}}$ is closed. 
Rewriting the wedge product ${\tens{p}\wedge\tens{q}}$ 
as the anti-symmetrization of their tensor product, 
we get 
\begin{equation}\label{wedgeCCKYT}
\begin{split}
 &\nabla_{\!e}(p_{[ab\dots}q_{cd\dots]}) = 
    (\nabla_{\!e} p_{[ab\dots})\, q_{cd\dots]}
    +p_{[ab\dots}\,(\nabla_{\!|e|}q_{cd\dots]})\\
 &\quad= g_{e[a}\bigl({\textstyle\frac{r}{D-r+1}}
 (\nabla_{|g|}p^g{}_{b\dots})\,q_{cd\dots]}\\
 &\mspace{140mu}   +(-1)^r \textstyle{\frac{s}{D-s+1}}p_{b\dots c}\,
 (\nabla_{|g|}q^g{}_{d\dots]})\bigr)\,,
\end{split}\raisetag{6.6ex}
\end{equation}\\
where we used the property \eqref{CCKYTdef} of ${\tens{p}}$ and ${\tens{q}}$. 
We can see by inspection that the result has the form
${g_{a[a_1}\alpha_{a_2\dots a_r]}}$, so, as we discussed after
Eq.~\eqref{dpdef}, ${\tp(\tens{p}\wedge\tens{q})=\tens{p}\wedge\tens{q}}$, and
hence ${\tens{p}\wedge\tens{q}}$ is a closed CKYT.

If we understand ${A^{(j)}_\mu}$ with ${\mu=1,\dots,n}$
and ${j=0,\dots,n-1}$ as an $n\!\!\times\!\! n$ matrix, its inverse is 
${B^{\,\mu}_{(j)}=(-x_\mu^2)^{n\!-\!1\!-\!j}/U_\mu}$
with ${U_\mu}$ defined in Eq.~\eqref{QXUcdef}. This means
\begin{equation}\label{Ainverse}
\sum_{j=0}^{n-1}(-x_\mu^2)^{n\!-\!1\!-\!j}\frac{A_\nu^{(j)}}{U_\mu}=\delta_\mu^\nu\;,\quad
\sum_{\mu=1}^{n} (-x_\mu^2)^{n\!-\!1\!-\!j} \frac{A_\mu^{(l)}}{U_\mu} = \delta_j^l\,.
\end{equation}
We also have its `extension' for ${j=n}$, ${l=0,\dots,n-1}$:
\begin{equation}\label{Aninverse}
\sum_{\mu=1}^n \frac{A_\mu^{(l)}}{x_\mu^2 U_\mu}  = \frac{A^{(l)}}{A^{(n)}}\;.
\end{equation}



\begin{thebibliography}{}
\bibitem{nt:EuclSign}
In the construction presented here we do not assume the specific form of the
metric; however, similar to the Kerr--NUT--AdS case \cite{CLP}, we assume that
the metric is written at least formally with Euclidean signature. The Wick
rotation to physical signature would bring some different signs in various
expressions, but the general structure would not be affected.
\bibitem{nt:velocity}
The velocity ${\tens{u}}$ appears here in a dual way. When proving that, e.g.,
${\tens{F}}$ is covariantly conserved along each geodesic, we understand
${\tens{u}}$ as the tangent vector with respect to an affine parameter  (not
necessary normalized). However, we can also use the velocity ${\tens{u}}$ in a
situation when we do not assume the whole trajectory to which ${\tens{u}}$ is
tangent. In such a case we refer to a point in the phase space which is
naturally realized as the cotangent bundle. The point in the phase space is
given by the point in the configuration space and by the momentum 1-form which
can be identified with the unnormalized velocity. Indeed, the geodetic motion
is generated by the Hamiltonian ${w=\tens{u}\cdot\tens{u}}$, the momentum is
${\tens{u}}$, and conserved quantities are those which Poisson commute with
${w}$.
\bibitem{nt:KYproof}
This was checked in \cite{KF} for ${D<10}$ and proved by the authors for
general dimensions $D$ by checking the condition \eqref{KYTdef} for the dual
${\tens{f}=*\tens{\cKY}}$, using a direct calculation of the connection
coefficients. Cf.\ also \cite{KNAcurv} for a recent independent calculation of
the connection coefficients.
\end{thebibliography}
\end{document}